\documentclass{elsart}

\usepackage{epsfig,amssymb}

\begin{document}

\begin{frontmatter}

\title{Effect of non--Gaussian noise sources in a noise induced
transition}

\author{Horacio S. Wio $^{1}$ and Ra\'ul Toral $^{2}$}
\address{Departament de F\'{\i}sica, Universitat de les Illes
Balears and \\ Instituto Mediterr\'aneo de Estudios Avanzados,
IMEDEA (CSIC-UIB), \\ Ed. Mateu Orfila, Campus UIB, E-07122 Palma
de Mallorca, Spain}

\thanks{E-mail: wio@imedea.uib.es. Permanent Address: Centro
At\'omico Bariloche (CNEA) and Instituto Balseiro (CNEA and
UNCuyo), 8400 San Carlos de Bariloche, Argentina.\\ $^2$ E-mail:
raul@imedea.uib.es.}


\begin{abstract}
Here we study a noise induced transition when the system is driven by a noise
source taken as colored and non-Gaussian. We show --using both, a theoretical
approximation and numerical simulations-- that there is a shift of the
transition as the noise departs from the Gaussian behavior. Also, we confirm
the reentrance effect found for colored Gaussian noise and show the behavior
of the transition line in the phase-like diagram as the noise departs from
Gaussianity in the large correlation time limit.
\end{abstract}


\end{frontmatter}


\section{INTRODUCTION}

During the last three decades a wealth of research results on
fluctuations or noise have lead us to the recognition that in many
situations noise can actually play a constructive role that
induces new ordering phenomena. Some examples are stochastic
resonance in zero--dimensional \cite{SRRMP} and extended systems
\cite{extend1,extend2}, noise induced transitions \cite{lefev},
noise induced phase transitions \cite{nipt1,nipt2}, noise induced
transport \cite{Ratch2,nipt3,nipt4}, noise sustained patterns
\cite{nsp}, etc.

Most of the studies on the noise induced phenomena indicated above
assume that the noise source has a Gaussian distribution (either
white or colored). In addition to the intrinsic interest in the
study of non--Gaussian noises, there are some experimental
evidences, particularly in sensory and biological systems
\cite{membr1}, indicating that at least in some of these phenomena
the noise sources could be non--Gaussian. The use of non--Gaussian
noises in studies on noise induced phenomena is scarce mainly due
to the mathematical difficulties. This is in contrast with the
existence of some analytical tools when working with Gaussian
(particularly white) noises.

Here we present some results on one of those noise induced
phenomena when driven by a noise source taken as colored and
non--Gaussian. It corresponds to a {\sl noise induced transition}
like those discussed in \cite{lefev}. The problem we discuss here
corresponds the so called {\sl genetic model} which follows from a
particular biological modeling or from a chemical reaction
\cite{lefev,a1}. In Ref. \cite{We0} the effect of a colored
Gaussian (Ornstein--Uhlenbeck) noise on that model was analyzed in
detail and a novel reentrance phenomenon in the phase diagram was
found. In order to study the effect that the non--Gaussian nature
of the noise can have in the transition, we have used here a
particular distribution \cite{borland1} whose departure
from the Gaussian behavior is governed by a parameter $q$: while
$q=1$ corresponds to Gaussian noise, the value $q>1$ produces a
long tail distribution and $q<1$ gives a cut-off distribution. As
discussed for other noise induced phenomena in \cite{We7}, we have
found that the departure from the Gaussian distribution for the
noise produces a strong effect. In this case such an effect
corresponds to an anticipation of the transition when the
distribution has long tails, and to a retardation when it is
cut-off. This effect could have interesting consequences in the
case of noise-induced-{\bf phase} transitions \cite{nipt1,nipt2}.

In the next section we briefly discuss the form and properties of
the non--Gaussian noise source. After that, we present the model
we analyze here and a simple analytical approximation to treat the
non--Gaussian noise that allows us to determine some of the
general features of the phase diagram. We continue discussing
numerical simulations that support and extend the theoretical
results. In the last section we draw some conclusions.

\section{NON--GAUSSIAN NOISE AND ITS PROPERTIES}

We start considering the following general form of a Langevin
equation
\begin{equation}
\dot{x} =f(x)+ g(x) \, \eta (t). \label{equis0}
\end{equation}
As described in the Introduction, we assume that the noise term
$\eta(t)$ has a non--Gaussian distribution. More precisely, we
consider that $\eta(t)$ is a Markovian process generated as the
solution of the following Langevin equation\cite{borland1}
\begin{equation}
\dot{\eta} =-\frac{1}{\tau }\frac{d}{d{\eta }}V_{q}(\eta )
+\frac{\sqrt{2D}}{\tau }\xi (t), \label{nu1}
\end{equation}
being $\xi (t)$ a standard Gaussian white noise of zero mean and
correlation $\langle\xi (t)\xi (t^{\prime })\rangle=\delta
(t-t^{\prime })$, and
\begin{equation}
\label{potential} \label{Vq1} V_{q}(\eta )=\frac{D}{\tau
(q-1)}\ln \left[1+\frac{\tau}{D}(q-1) \frac{\eta^{2}}{2}\right] .
\end{equation}
Although we believe that our results are quite general, such a
particular form for the noise $\eta(t)$ allows us to easily
control the departure from the Gaussian behavior by changing a
single parameter $q$. $D$ and $\tau$ are noise parameters related
to the noise intensity and the correlation time, as we now detail.
The stationary properties of the noise $\eta$, including the time
correlation function, have been studied in \cite{We5} and here we
summarize the main results. The stationary probability
distribution is given by
\begin{equation}\label{pdf}
P_{q}^{st}(\eta)=\frac{1}{Z_{q}}\left[ 1+\frac{\tau}{D} (q-1)
 \frac{\eta^{2}}{2}\right] ^{\frac{-1}{q-1}},
\end{equation}
where $Z_{q}$ is the normalization factor. This distribution can
be normalized only for $q < 3$. The first moment, $\langle \eta
\rangle=0$, is always equal to zero, and the second moment,
\begin{equation}
\langle\eta^2\rangle=\frac{2D}{\tau (5-3q)}, \label{2nd}
\end{equation}
is finite only for $q<5/3$. Furthermore, for $q<1$, the
distribution has a cut-off and it is only defined for $|\eta|<
\eta_c\equiv \sqrt{\frac{2D} {\tau (1-q)}}$. Finally, the
correlation time $\tau _q$ of the stationary regime of the process
$\eta(t)$ diverges near $q=5/3$ and it can be approximated over
the whole range of values of $q$ as $\tau _q\approx 2\tau
/(5-3q)$. Clearly, when $q \to 1$ we recover the limit of $\eta$
being a Gaussian colored noise, namely the Ornstein--Uhlenbeck
process, $\xi_{ou}(t)$, with correlations $\langle
\xi_{ou}(t)\xi_{ou}(t')\rangle=\frac{D}{\tau}{\rm
e}^{-|t-t'|/\tau}$ and probability distribution
$P^{st}(\xi_{ou})=Z^{-1}{\rm e}^{-\frac{\tau}{2D}\xi_{ou}^2}$.

In Ref. \cite{We5}, an effective Markovian approximation to the
process $\eta$ via a path integral procedure was obtained, that is
analogous to the ``unified colored noise approximation" obtained
in \cite{ee} for the case of Ornstein--Uhlenbeck noise. Such an
approximation allowed us to get quasi-analytical results for the
mean-first-passage time or transition rate. These results and its
dependence on the different parameters in the case of a double
well potential were compared with extensive numerical simulations
with excellent agreement. That approximation was also exploited to
study stochastic resonance with theoretical results \cite{We4}
that are in very good agreement with experimental ones
\cite{WeEx}.

\section{$q \approx 1$ APPROXIMATION FOR THE NON--GAUSSIAN NOISE}

The ``effective Markovian approximation" introduced in Ref.
\cite{We5} and indicated above has been shown to have some drawbacks.
As discussed in \cite{We5}, such an approximation renders a non
normalizable stationary probability distribution function,
$P_q^{st}(\eta)$. This seems to be a general characteristic of
such an approximation that, at least in the case discussed in
\cite{We5}, was solved introducing an adequate and controlled
cutoff. However, it is not clear that such a methodology could be
always used, and particularly for the models we can analyze in
relation to noise induced transitions. Hence, for the kind of
models we use in order to study the effect of non--Gaussian noises
on a noise induced transition we have resorted to a more simple,
albeit physically reasonable, approximation. Our aim is at least
to obtain some of the main trends in the region $|q - 1|\ll 1$
(both for $q < 1$ and $q>1$). The idea is that in such a region
the non--Gaussian noise will only slightly depart from the
Gaussian behavior. Hence in Eq. (\ref{nu1}) we adopt
\begin{eqnarray}
\label{Vq2} \frac{1}{\tau }\frac{\partial V_{q}(\eta )}{\partial
\eta } & = & \frac{\eta }{\tau}{\displaystyle \left[
1+\frac{\tau}{D}(q-1) \frac{ \eta^{2} }{2} \right]^{-1}} \\
\nonumber & \approx & \frac{\eta }{\tau }{\displaystyle \left[
1+\frac{\tau}{D}(q-1) \frac{\langle \eta^{2}\rangle }{2}
\right]^{-1}}\,\, \equiv \,\, a(\tau ) \eta ,
\end{eqnarray}
with $\displaystyle a(\tau) = \frac{5 - 3 q}{2 \tau (2 - q)}$.
This implies a kind of renormalized Ornstein--Uhlenbeck (Gaussian)
noise. Using this form of the noise we proceed to study the
problem in Eq.(\ref{equis0}).

The {\sl genetic model} is described by the following equation
\cite{lefev,a1,We0}
\begin{equation}
\label{genic} \dot x=\frac 12-x+\lambda x(1-x)+x(1-x) \eta (t),
\label{eq:5}
\end{equation}
and, according to the discussion above, we consider in this work
that $\eta (t)$ is a non--Gaussian noise described by
Eqs.(\ref{nu1}-\ref{potential}).

The analytical treatment simplifies if we make the change of
variables
\begin{equation}
y = \ln \Bigl( \frac{x}{1-x} \Bigr), \label{eq:6}
\end{equation}
so that the original multiplicative stochastic differential
equation (\ref{eq:5}) transforms into a new one but now with an
additive noise
\begin{equation}
\dot y = - \sinh (y) + \lambda + \eta (t). \label{eq:7}
\end{equation}
From now on, as in Ref.\cite{We0}, and in order to simplify the
algebra, in what follows we adopt $\lambda = 0$.

The use of the approximation Eq. (\ref{Vq2})
(that we reiterate could be valid only for $|q-1| \ll 1$)
together with the ``unified colored noise approximation"
\cite{ee}, yields
\begin{equation}
\dot y \approx f(y) \frac{1 }{[1 - \frac{f'(y)}{a(\tau)}]} +
\frac{1}{\tau a(\tau)} \frac{1}{[1 - \frac{f'(y)}{a(\tau)}]} \xi
(t),  \label{eq:8}
\end{equation}
with $f(y) = - \sinh (y)$. This result can be obtained in two
ways: (a) Applying a direct adiabatic-like elimination procedure
to the Langevin system given by Eq. (\ref{eq:7}) and Eqs.
(\ref{nu1},\ref{Vq2}) \cite{ee}. That is, taking the derivative of Eq.
(\ref{eq:7}) with respect to $t$ and setting $\ddot{y}=0$. (b) A
formal one, through the application of the path-integral formalism
to the indicated non Markovian Langevin equations, supplemented
with a consistent Markovian approximation scheme \cite{We9}.

Since, by the use of these ``drastic" approximations, the problem
has been reduced to Eq. (\ref{eq:8}) which contains a white,
Gaussian, noise, it is easy now, using standard techniques, to
obtain the  stationary probability density function. For the
original variable $x$, it reads
\begin{equation}
P_{st}(x) = N \exp[F(x)],  \label{eq:10}
\end{equation}
where $N$ is a normalization constant and
\begin{eqnarray}
F(x) = &-& \frac{1}{8 D} \left( \frac{5-3q}{2-q} \right)^{2}
\left( \frac{1-2x+2x^{2}}{x(1-x)} \right) \\
\nonumber & & \,\,\,\,\,\,\,\,\,\,\,\,  \times \left( 1 + \left(
\frac{2-q}{5-3q} \right) \left( \frac{\tau}{2} \right)
\left( \frac{1-2x+2x^{2}}{x (1-x)} \right) \right)\\
\nonumber
 &+& \ln \left( 1 +  \left(\frac{2-q}{5-3q} \right) \tau
 \left(\frac{1-2x+2x^{2}}{x (1-x)} \right) \right) - \ln(x(1-x)).
\label{eq:11}
\end{eqnarray}

\begin{figure}[!ht]
\centerline{\makebox{\epsfig{figure=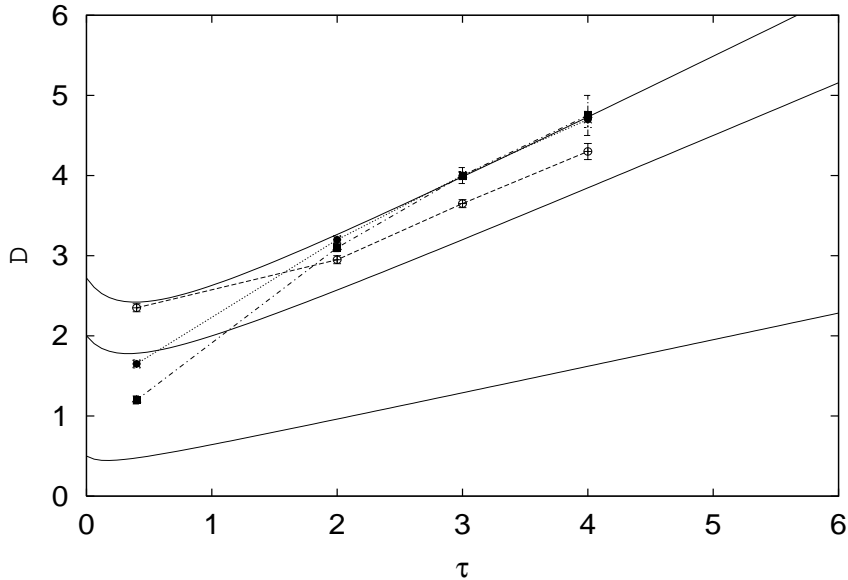,width=8cm,height=12cm,angle=-90}}}
\vspace{0.5truecm}\caption{\label{fig1} Phase diagram $D$ vs.
$\tau$ for different values of $q$. Probability distributions are
bimodal for points above the curves and unimodal below. Solid
lines are the theoretical predictions, as given by Eq.
(\ref{eq:12}) for $q=0.5,\,q=1,\,q=1.5$ (from bottom to top). The
numerical results are indicated by symbols joined by a dashed line
for $q=0.5$, a dotted line for $q=1$, and a dash-dotted for
$q=1.5$. It is clear the reentrance effect and the shift of the
transition line. }
\end{figure}

A noise induced transition\cite{lefev} appears when this density
function changes from being unimodal to bimodal. More
specifically, the point $x=1/2$ changes from being a maximum to
being a minimum. The condition of extremum at $x=1/2$ and the
condition for finding the boundary separating the regions where
$x=1/2$ is a maximum or a minimum yield
\begin{equation}
D(\tau) = \frac{1}{2} \left( \frac{5-3q}{2-q} \right)^{2} \left( 1
+ \frac{2-q}{5-3q} 2 \tau \right)^{2} \left( 1 + \frac{2-q}{5-3q}
6 \tau \right)^{-1}.  \label{eq:12}
\end{equation}
such that for a fixed $\tau$, the distribution is bimodal for
$D>D(\tau)$ and it is unimodal for $D<D(\tau)$. In the limit $q
\rightarrow 1$, this expression reduces to the result found in
\cite{We0} for  $q = 1$:
\begin{equation}
D_{0}(\tau) = 2 \frac{ [1 +  \tau ]^{2}}{1 + 3 \tau}.
\label{eq:13}
\end{equation}
Figure \ref{fig1} plots the $D(\tau)$ boundary (the ``transition
line") for different values of $q$ in the $(D,\tau)$ plane.
Several features can be noticed in this figure. It is clear that a
minimum value of the noise intensity $D$ is always needed to
induce the transition. The dependence with the parameter $q$
appears as an advancement of the transition to smaller values of
$D$ when $q > 1$ and a retardation when $q < 1$. We believe that
this will be a general behavior for long tail (power--law) and
cut--off noise distributions, respectively. A closer look shows
the persistence of the reentrance effect previously found in
\cite{We0}: for some fixed values of the noise intensity  $D$ it
is possible to obtain a transition, first from unimodal to bimodal
and then back to unimodal, just by varying the time correlation
$\tau$ of the noise.

These main features, as obtained from our very simplifying
assumptions are found to be in qualitative agreement with the
results coming from numerical simulations of the (not simplified)
set of equations (\ref{nu1},\ref{potential},\ref{genic}) (details
of the numerical integration scheme are given in the appendix). In
figure \ref{fig2} we show that the probability distribution can
switch from unimodal to bimodal just by increasing the parameter
$q$, for fixed values of $D$ and $\tau $. The reentrance effect is
illustrated, for $q=1$ in figure \ref{fig3}.

The numerical simulations show that there is a crossing of the
transition lines for $q<1$ and $q \geq 1$, an aspect which is not
reproduced by the approximated theoretical results. Another aspect
observed in figure \ref{fig1} is that for large $\tau $, the
theoretically predicted transition line, has a smaller linear
slope for increasing values of $q$. In fact, according to Eq.
(\ref{eq:12}) we have $D\sim \frac{5-3q}{3(2-q)}\tau$ in that
limit. There seems to be not such a strong $q$ dependence of the
slope of the curves for large $\tau$ in the numerical results.
Otherwise, the numerical and theoretical results are in good
qualitative agreement.

\begin{figure}[!ht]
\centerline{\makebox{\epsfig{figure=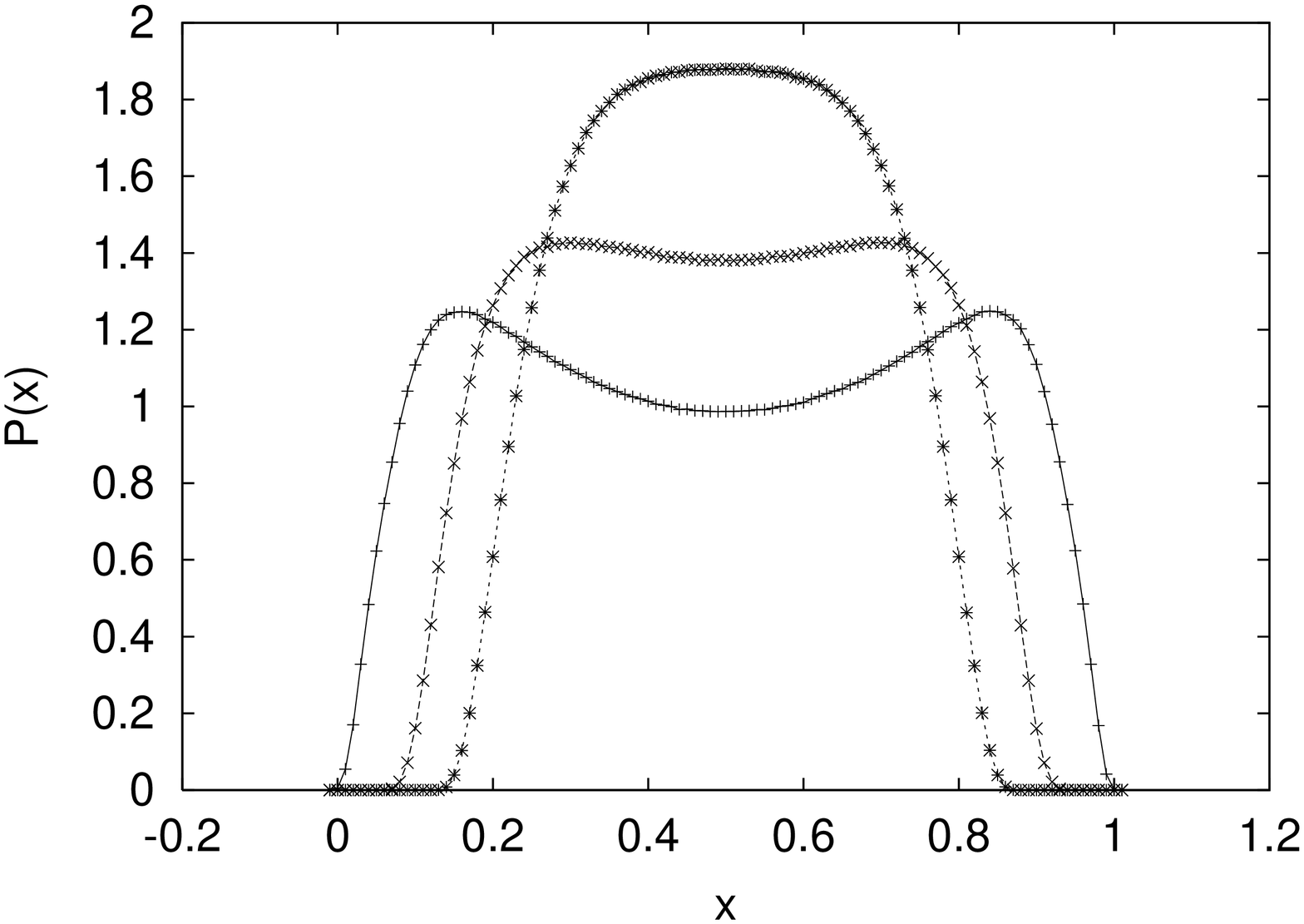,width=8cm,height=12cm,angle=-90}}}
\vspace{0.5truecm}\caption{\label{fig2} Simulation results for the
stationary probability distribution $P_{st}(x)$ for fixed values
$D=2$, $\tau =0.4$ and different values of $q$: $q=0.5$ (*), $q=1$
(x), $q=1.5$ (+). It is clear that it is possible to induce a
transition from unimodal to bimodal just by changing the parameter
$q$,  measuring the departure from the Gaussian distribution for
the noise source.}
\end{figure}

\begin{figure}[!ht]
\centerline{\makebox{\epsfig{figure=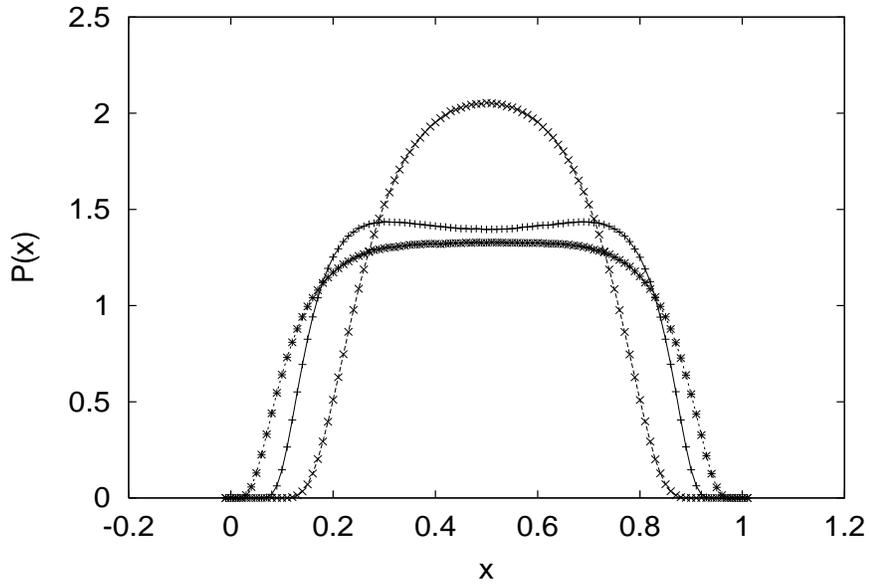,width=8cm,height=12cm,angle=-90}}}
\vspace{0.5truecm}\caption{\label{fig3} Simulation results for the
stationary probability distribution $P_{st}(x)$ for $q=1$
(Gaussian case) for fixed $D=1.95$, and different values of $\tau
$: $\tau =0.$ (x), $\tau =0.4$ (+), $\tau =1.0$ (*). The
reentrance effect indicated in the phase diagram $D$ vs. $\tau $
is apparent..}
\end{figure}

\section{FINAL REMARKS}

We have presented some preliminary results on a noise induced
phenomenon driven by a colored and non--Gaussian noise source that
was generated by a $q$-distribution. The phenomenon studied here
is a {\sl noise induced transition} like those discussed in
\cite{lefev}, and corresponds the so called {\sl genetic model}.
We have resorted to a simple approximation --that avoids the
difficulties of the effective Markovian approximation in
\cite{We5}-- introduced with the idea that for $|q - 1| \ll 1$,
the non--Gaussian noise will only slightly depart from the
Gaussian one, and its effect being only a correction to the
Gaussian behavior. The resulting Eq. (\ref{eq:12}) shows some of
the right tendencies with $q> 1$ or $q<1$, although some features
observed in the simulations (crossing of the transition lines and
slight dependence of the linear slope for large $\tau$) are not
reproduced by the simplified theoretical treatment. In addition,
the stationary probability density function defined in
Eqs.(\ref{eq:10}) and (\ref{eq:11}) clearly shows, for constant
values of $D$ and $\tau $ and as a function of $q$, the transition
from unimodal to bimodal behavior. Again, these general results
and tendencies, together with the reentrance effect, have been
confirmed by extensive numerical simulations as shown in figures
1, 2 and 3. However, the agreement between theory and simulations
is only qualitative, as the first are unable to reproduce some of
the details revealed by the second. For instance, from the
numerical simulations a more complex behavior of the transition
line for large $\tau $ is apparent, while it is not present in the
theoretical predictions (see Fig. 1).

These results are in the same line that a series of previous
studies on noise induced phenomena where the noise source was
non--Gaussian. In all the cases, we have found that the response
of the system results to be enhanced or to have a strong
dependence for values of the parameter $q$ corresponding to a
departure from Gaussian behavior. Some of such studies, including
stochastic resonance \cite{We4,WeEx}, gated traps \cite{We3},
Brownian motors \cite{WeR}, have been reviewed in \cite{We7}.

Even though our results are so far only preliminary and further
studies are required, we expect that the phenomenon discussed here
will have remarkable effects on the noise induced phase
transitions \cite{nipt2} as well as in the problem of coupled
ratchets \cite{nipt3,nipt4}. For instance, we can expect changes
in the phase diagram of the first, or in the hysteresis cycle of
the second, that could have interesting consequences in
technological applications. Such studies will be the subject of
further work.

\section*{ACKNOWLEDGMENTS} This works is supported by MCyT (Spain)
and FEDER, projects BFM2001-0341-C02-01 and BFM2000-1108. HSW
acknowledges the partial support from CONICET and ANPCyT, both
Argentine agencies, and thanks the MECyD, Spain, for an award
within the {\it Programa de Sab\'aticos para Profesores
Visitantes}, and the Universitat de les Illes Balears for the kind
hospitality extended to him.

\section*{APPENDIX: NUMERICAL SIMULATIONS}

In order to test our theoretical predictions,we have resorted to
numerical simulations of the system indicated by Eqs.
(\ref{genic}) and (\ref{nu1},\ref{potential}). For the numerical
integration of those equations we have used a second order
stochastic Runge--Kutta type algorithm \cite{raul}. For the
general problem
\begin{eqnarray}
\label{dotx} \dot x & = & f(x)+g(x)\eta(t)\cr \dot \eta & = &
q(\eta)+\epsilon\xi(t),
\end{eqnarray}
where $\xi(t)$ is a white noise of mean equal to zero and
correlations $\langle \xi(t)\xi(t')\rangle =\delta(t-t')$, the
algorithm is as following: time is discretized in steps of size
$h$ and the value of the process $\eta(t)$ at the discretized
times is obtained by the recursion relation
\begin{eqnarray}
k(t) & = & h q(\eta(t))\cr l(t) & = & \epsilon h^{1/2}G(t)\cr
\eta(t+h)& =&
\eta(t)+\frac{h}{2}\left[q(\eta(t))+q(\eta(t)+k(t)+l(t))
\right]+l(t),
\end{eqnarray}
with the initial value $\eta(0)=0$. Here the numbers $G(t)$ for
different times $t$ are independent Gaussian random numbers of
mean equal to zero and variance equal to one. They have been
obtained by using a particularly efficient generator\cite{tc93}.
In the case $q<1$, it might occur that the generated value of
$G(t)$ is such that $\eta(t+ h)$ is outside the cut-off interval
$(-\eta_c,\eta_c)$. If this happens, that value is discarded and a
new $G(t)$ is generated until the condition $|\eta(t+h)|<\eta_c$
is fulfilled. We have found that the percentage of discarded
$G(t)$ values is less than $5\times 10^{-4}$ in the worst case
analyzed here corresponding to $q=0.5,\,\tau=4$. No such a problem
ever occurs for $q\ge 1$.

Once the values of $\eta(t)$ at the discretized times have been
obtained, the integration of (\ref{dotx}) proceeds with a standard
Runge--Kutta second order method:
\begin{eqnarray}
k_x(t) & = & f(x(t)) \cr
x(t+h) & = & x(t)+\frac{h}{2} [ f(x(t))+ f(x(t)+k_x(t))+\cr
& &  \hspace{1.7truecm}g(x(t))\eta(t)+g(x(t)+k_x(t))\eta(t+h)].
\end{eqnarray}
We have used a time step $h=0.01$ and the histogram for
$P^{st}(x)$ has been obtained from $10^8$ values of $x$ separated
by a time $t=1$ (corresponding, hence, to $100$ integration
steps).

\end{document}